\documentclass[conference]{IEEEtran}
\IEEEoverridecommandlockouts
\usepackage{cite}
\usepackage{amsmath,amssymb,amsfonts}
\usepackage{algorithmic}
\usepackage{graphicx}
\usepackage{textcomp}
\usepackage{xcolor}
\usepackage{enumitem}
\usepackage{url}
\def\BibTeX{{\rm B\kern-.05em{\sc i\kern-.025em b}\kern-.08em
    T\kern-.1667em\lower.7ex\hbox{E}\kern-.125emX}}
\begin{document}

\title{Comparing Normalization Methods for Portfolio Optimization with Reinforcement Learning\\
\thanks{The authors would like to thank the \textit{Programa de Bolsas Itaú} (PBI) of the \textit{Centro de Ciência de Dados} (C$^2$D) at Escola Politécnica at USP, supported by \textit{Itaú Unibanco S.A.}, and the Brazilian National Council for Scientific and Technological Development (CNPq Grant N. 310085/2020-9).}
}

\author{\IEEEauthorblockN{Caio de Souza Barbosa Costa}
\IEEEauthorblockA{\textit{Escola Politécnica} \\
\textit{Universidade de São Paulo (USP)}\\
São Paulo, Brazil \\
0009-0009-7780-7009}
\and
\IEEEauthorblockN{Anna Helena Reali Costa}
\IEEEauthorblockA{\textit{Escola Politécnica} \\
\textit{Universidade de São Paulo (USP)}\\
São Paulo, Brazil \\
0000-0001-7309-4528}
}


\maketitle

\begin{abstract}
Recently, reinforcement learning has achieved remarkable results in various domains, including robotics, games, natural language processing, and finance. In the financial domain, this approach has been applied to tasks such as portfolio optimization, where an agent continuously adjusts the allocation of assets within a financial portfolio to maximize profit. Numerous studies have introduced new simulation environments, neural network architectures, and training algorithms for this purpose. Among these, a domain-specific policy gradient algorithm has gained significant attention in the research community for being lightweight, fast, and for outperforming other approaches. However, recent studies have shown that this algorithm can yield inconsistent results and underperform, especially when the portfolio does not consist of cryptocurrencies. One possible explanation for this issue is that the commonly used state normalization method may cause the agent to lose critical information about the true value of the assets being traded. This paper explores this hypothesis by evaluating two of the most widely used normalization methods across three different markets (IBOVESPA, NYSE, and cryptocurrencies) and comparing them with the standard practice of normalizing data before training. The results indicate that, in this specific domain, the state normalization can indeed degrade the agent's performance.
\end{abstract}


\section{Introduction}
Portfolio optimization is a task in which an agent constantly defines the percentage of each asset in a financial portfolio considering market trends that can be identified through quantitative analyses, political and financial news, public opinion, among others \cite{Gunjan_2022_BriefReviewPortfolio}. With the advent of algorithmic trading, computational agents have been increasingly utilized to perform such task, and, thanks to the emergence of Machine Learning techniques, several supervised approaches have been applied to forecast the future value of assets in the portfolio and assist the decision-making \cite{Henrique_2019_LiteratureReviewMachine}. Those approaches, however, are very sensitive to real-world features of the financial market, such as brokerage fees \cite{Deng_2017_DeepDirectReinforcement}, and, thus, a technique that maximizes the future return of the portfolio considering these characteristics is more appropriate.

In this context, the Reinforcement Learning (RL) technique is a good fit: the agent repeatedly interacts with the environment and learns the optimal behavior by trial and error, being guided by a reward function, which is more positive when the agent's actions produce better results \cite{Sutton_2018_ReinforcementLearningIntroduction}. If the reward function is related to the profit made by the agent after it performs its actions, the RL formulation conveniently adheres to the calculation of the future return of the financial portfolio and it can be optimized with the usage of several well-known training algorithms \cite{Felizardo_2022_ReinforcementLearningApplied}.

This idea was applied in \cite{Jiang_2017_DeepReinforcementLearning}, which introduced a framework to optimize financial portfolios composed of cryptocurrencies and created a realistic formulation that simulated the effects of the market considering transaction costs. By using a convolutional neural network called \textit{Ensemble of Identical Independent Evaluators} (EIIE) to process the time series of the price of the cryptocurrencies and developing a domain-specific policy gradient algorithm, the framework outperformed several other classical portfolio strategies. 

The approach utilized in \cite{Jiang_2017_DeepReinforcementLearning} had a huge impact in the research field and was the foundation of several other works. \cite{Shi_2019_MultiScaleTemporalFeature} changed the Deep Learning architecture and applied a multi-scale convolution inspired by \textit{Xception} networks \cite{Chollet_2017_XceptionDeepLearning} to create the \textit{Ensemble of Identical Independent Inception} (EI$^3$) architecture, which outperformed EIIE. \cite{Soleymani_2021_DeepGraphConvolutional} and \cite{Shi_2022_GPMGraphConvolutional} went further, and developed agents that made use of graph neural networks \cite{Wu_2021_ComprehensiveSurveyGraph} to model the relationship between assets in the portfolio, introducing, respectively, \textit{DeepPocket} and \textit{Graph-based Portfolio Management} (GPM), which generates more profit than the convolutional solutions. \cite{Xu_2020_RelationAwareTransformerPortfolio}, on the other hand, applied Transformer architectures \cite{Vaswani_2023_AttentionAllYou} to the framework in order to account for the non-Euclidean characteristics of the temporal series of the assets and also achieved great results.

The large adoption of the framework introduced in \cite{Jiang_2017_DeepReinforcementLearning} is a result of the fact that the domain-specific training algorithm converges faster and leads to better policies \cite{Liang_2018_AdversarialDeepReinforcement} than other state-of-the-art algorithms such as \textit{Deep Deterministic Policy Gradient} (DDPG) \cite{Lillicrap_2015_ContinuousControlDeep} and \textit{Proximal Policy Optimization} (PPO) \cite{Schulman_2017_ProximalPolicyOptimization}. However, recent studies point that the framework is not as reliable as it appears in stock markets: its performance is considerably worse in comparison to the cryptocurrency market \cite{Li_2024_DeepReinforcementLearning}. One of the reasons that can be generating this disparity is the state regularization method utilized, whose effectiveness was not investigated by the scientific community in this specific use-case.

To mitigate this issue, this paper compares the performance of three normalization methods that can be used in portfolio optimization agents trained with reinforcement learning. Two of them are state normalizations commonly used in the field: the normalization by the last closing price and the normalization by the last value. The other technique consists of normalizing the time series of assets independently before the training process. This work aims to answer the following research questions:

\begin{enumerate}
    \item Using data normalization instead of state normalization improves the performance of the agent in stock markets?
    \item What is the effect of applying data normalization in cryptocurrency portfolios such as the one introduced in \cite{Jiang_2017_DeepReinforcementLearning}?
\end{enumerate}

This paper is organized as follows. Section \ref{sec:formulation} introduces the mathematical formulation of the portfolio optimization task. Section \ref{sec:reinforcement-learning} explains what is reinforcement learning and how it is possible to apply it to the problem in case. Section \ref{sec:normalization} describes the importance of normalization methods in the task and lists a few possible techniques that can be used. Sections \ref{sec:experiments} and \ref{sec:results} discuss, respectively, the experiments that are conducted to test the normalization techniques and its results. Finally, Section \ref{sec:conclusion} concludes this document.

\section{Mathematical Formulation of Portfolio Optimization}
\label{sec:formulation}

In the portfolio optimization task, at each time step $t$ and for a portfolio of $n$ assets, the agent is responsible for defining the \textit{portfolio vector} or \textit{weights vector} $\vec{W_{t}} \in \mathbb{R}^{n+1}$ whose elements are the rate (or weight) of each asset in the portfolio composition. Considering $\vec{W_{t, i}}$ the $i$-th element of the vector, it is adopted that $\vec{W_{t, 0}}$ refers to the uninvested money while the other elements are related to the assets of the portfolio. Therefore, the following conditions must be met:
\begin{align}
\label{eq:constraints}
    0 \le \vec{W_{t, i}} \le 1, &&
    \sum\limits_{i=0}^{n} \vec{W_{t, i}} = 1.
\end{align}

At each time step $t$, there is also a price vector $\vec{P_{t}} \in \mathbb{R}^{n+1}$ whose elements are the prices of every component in the portfolio. Like in $\vec{W_{t}}$, it is considered that $\vec{P_{t, 0}}$ refers to the remaining cash of the portfolio and, thus, $\vec{P_{t, 0}} = 1$ and $\vec{P_{t, i}}$, in which $i \in \{1, 2, ..., n\}$, are calculated with respect to $\vec{P_{t, 0}}$.

During every simulation step, the price of the assets in the portfolio change from $\vec{P_{t}}$ to $\vec{P_{t+1}}$, changing the portfolio vector from $\vec{W_{t}}$ to $\vec{W_{t}^f}$ through the following equation:
\begin{equation}
    \vec{W_{t}^{f}} = \frac{(\vec{P_{t}} \oslash \vec{P_{t-1}}) \odot \vec{W_{t}}}{(\vec{P_{t}} \oslash \vec{P_{t-1}}) \cdot \vec{W_{t}}},
\end{equation}
in which $\oslash$ denotes the element-wise division, $\odot$ is the element-wise multiplication and $\cdot$ represents the dot product of vectors.

In the beginning of step $t$, the value of the portfolio is defined as $V_{t}$, but, just like portfolio vector, it changes to $V_{t}^f$ due to the variation of the prices of the assets. The equation to calculate the new value is given by:
\begin{equation}
\label{eq:final-value}
    V_{t}^{f} = V_{t} \Bigl(\vec{W_{t}} \cdot (\vec{P_{t}} \oslash \vec{P_{t-1}})\Bigl).
\end{equation}

In the next time step, a new portfolio vector $\vec{W_{t+1}}$ is set by the agent and the value of the portfolio changes:
\begin{equation}
\label{eq:next-value}
    V_{t+1} = \mu_{t+1} V_{t}^{f},
\end{equation}
in which $\mu_{t+1} \in [0, 1]$ is a factor that reduces the portfolio value simulating the effects of the transaction costs from the rebalancing ($\vec{W_{t}^{f}} \rightarrow \vec{W_{t+1}}$). Details about the calculation of $\mu_{t+1}$ can be found in \cite{Jiang_2017_DeepReinforcementLearning}. 

The simulation, then, continues, following the same equations presented before until the simulation period is finished. In the initial conditions, no money is invested in the assets, so that $\vec{W_{0}} = [1, 0, 0, ..., 0]$ and the agent aims to discover the sequence of portfolio vectors $[\vec{W_{1}}, \vec{W_{2}}, ..., \vec{W_{T}}]$ that maximizes the final value of the portfolio $V_{T}^{f}$, in which $T$ is the final time step of the simulation being run.

Note that, in this formulation, it is considered that the assets can be immediately bought and sold for the last closing price and that the agent's actions do not influence the market. These are good approximations when the portfolio is composed of high-volume assets and when its initial value is relatively small, considering the whole money available in the trading market.

\section{Reinforcement Learning}
\label{sec:reinforcement-learning}

\begin{figure}[!ht]
    \centering
    \includegraphics[width=0.9\linewidth]{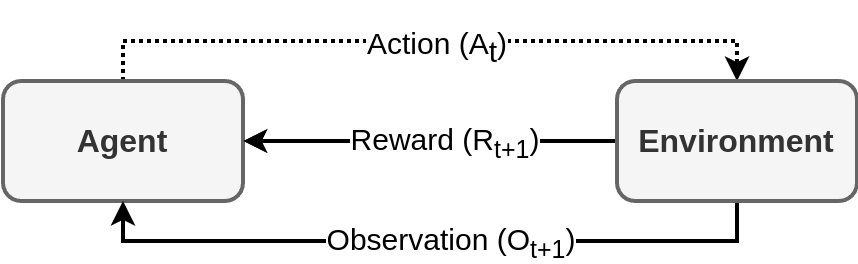}
    \caption{A cycle of reinforcement learning.}
    \label{fig:reinforcement-learning}
\end{figure}

The Reinforcement Learning is a technique in which an agent interacts with an environment that provides, at each time step $t$, observations $O_{t}$ with data to be used in the decision-making. The agent, then, defines a state $S_{t}$, which summarizes its current situation and is utilized by the agent to choose an action $A_{t}$ to be performed in the environment. After that, the environment transitions to another state and returns a new observation $O_{t+1}$ and a reward $R_{t+1}$ which is a numerical value that is bigger when the action performed leads to a favorable outcome. The process continues until a terminal state is achieved (or indefinitely, if such a state does not exist) and the agent's objective is to maximize the expected return of the rewards. Therefore, the agent learns continuously interacting with the environment by trial and error. Figure \ref{fig:reinforcement-learning} summarizes the interactions between the agent and the environment during a reinforcement learning training cycle.

In order to transform a state $S_{t}$ into an action $A_{t}$, the agent utilizes a policy of actions, which is a function $\pi: S \rightarrow A$ that maps all the possible states to all the possible actions. Due to the complexity of real-world problems, in which states and actions can usually be represented by multidimensional vectors of continuous values, neural networks \cite{Haykin_1999_NeuralNetworksComprehensive} are commonly used as the policy of the agent since it can approximate close values without the need to divide the state space or the action space in small intervals of defined size \cite{Sutton_2018_ReinforcementLearningIntroduction}. The policy that provides the agent with the maximum expected return (and thus, the policy that the agent is trying to learn) is called the \textit{optimum policy}.

\subsection{Applying to Portfolio Optimization}

In order to apply the RL approach to the portfolio optimization task, it is necessary to define the states, actions, rewards and the policy that the agent will use.

\begin{description}
    \item[State:] In this work, the agent will utilize the time series of the price of the assets in the portfolio to define the market situation. Therefore, for a portfolio composed of $n$ assets, it is possible to represent all the time series with a matrix of shape $(n, t)$, in which $t$ is the size of the time series, which is commonly called \textit{time window}. However, recent research demonstrated that it is beneficial to utilize more features in the state, in particular, the closing, high and low prices of the assets in the portfolio \cite{Weng_2020_PortfolioTradingSystem}. To be able to represent several features, the state is composed of a three-dimensional matrix of shape $(f, n, t)$, in which $f$ is the number of features considered. A visualization of the utilized state is presented in Figure \ref{fig:state-space}.
    \begin{figure}[!ht]
        \centering
        \includegraphics[width=0.5\linewidth]{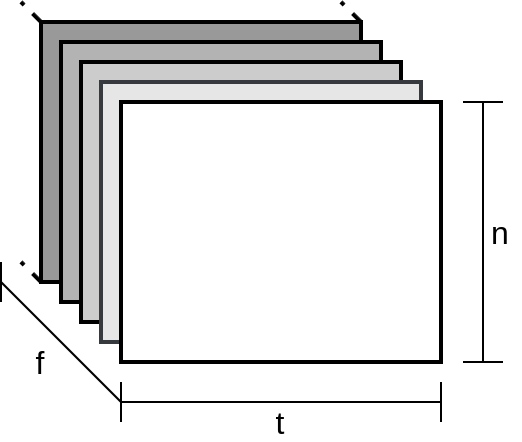}
        \caption{Shape of the state of the agent.}
        \label{fig:state-space}
    \end{figure}
    \item[Action:] The action of the agent is the portfolio vector introduced in Section \ref{sec:formulation}.
    \item[Reward:] The reward at time step $t$ is the logarithmic rate of return. It is positive when the value of the portfolio increases and negative otherwise. It can be calculated with the following equation:
    \begin{equation}
        r_{t} = ln \Bigl(\frac{V_{t}^{f}}{V_{t-1}^{f}}\Bigl).
    \end{equation}
    \item[Policy:] For all the experiments of this work, the EIIE architecture introduced in \cite{Jiang_2017_DeepReinforcementLearning} is utilized, since it is fast and is a common baseline in the research field. This architecture make use of convolutional filters in order to independently process the time series of the assets using the same model. An interesting feature of the EIIE policy is that, in order to be able to infer the effects of the transaction costs in the decision-making and improve the performance, it also considers the last performed action, so that $A_{t} = \pi (S_{t}, A_{t-1})$.
\end{description}

\subsection{Policy Gradient for Portfolio Optimization}
\label{subsec:policy-gradient}

Several RL training algorithms can be utilized to train a portfolio optimization agent, but, according to \cite{Liang_2018_AdversarialDeepReinforcement}, the domain-specific policy gradient (PG) algorithm introduced in \cite{Jiang_2017_DeepReinforcementLearning} achieves better performance in comparison to other general algorithms. 

In the PG algorithm, the agent interacts with the environment and saves its experiences in a replay buffer so that it contains one experience for each time step $t$ in the simulation episode (an episode can be defined as the set of simulation steps needed to cover the training period defined by the input historical data). When the buffer is completely filled, the agent starts to sample sequential batches of data: being $\Delta t$ the batch size chosen, a sequential batch is composed of experiences from the period $[T, T + \Delta t]$, in which $T$ is the first time step of the batch.

The batches of data are, then, utilized to update the parameters $\theta$ of the agent's policy $\pi$ in order to maximize the objective function given by:
\begin{equation}
\label{eq:objective-function}
    \sum\limits_{t=T_{1}}^{T_{2}} \frac{ln(\mu_{t}(\vec{W_{t}} \cdot (\vec{P_{t}} \oslash \vec{P_{t-1}}))}{T_{2} - T_{1} + 1},
\end{equation}
in which $T_{1}$ is the initial time step of the batch used and $T_{2}$ is the last one. Note that this objective function is directly obtained by equations \ref{eq:final-value} and \ref{eq:next-value}, so that the agents policy is optimized to maximize the profit in the batch of experiences.

Since the price variation of the assets in the training data does not change, the PG algorithm directly updates the experiences of the replay buffer after the optimization, without the need to be constantly interacting with the environment. This characteristic of the algorithm considerably accelerates the training process.

Finally, it is important to highlight that the agent is able to perform \textit{online learning}, so that new experiences can be added to the replay buffer and used by the agent to adapt itself to unknwon data. This is one of the biggest advantages in comparison to supervised approaches \cite{Deng_2017_DeepDirectReinforcement}, since the agent can be continuously trained while still on production.

\section{Normalization Methods and its Advantages}
\label{sec:normalization}

Normalizing the inputs of a deep neural network is an essential step to improve its performance \cite{Singh_2020_InvestigatingImpactData}: using normalized data prevents that numerical inputs with different scales degrades the learning process. In the portfolio optimization task, the normalization methods utilized in the research can be classified in two categories.

\subsection{State Normalization}
\begin{figure}[!ht]
    \centering
    \includegraphics[width=0.55\linewidth]{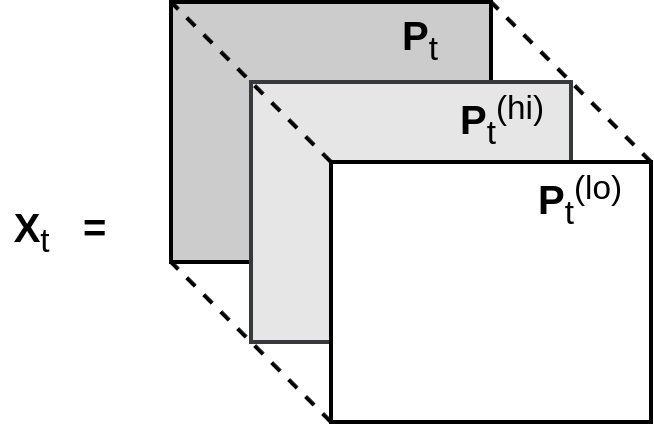}
    \caption{Construction of state $\vec{X_{t}}$ with $\vec{P_{t}}$, $\vec{P_{t}}^{(hi)}$ and $\vec{P_{t}}^{(lo)}$.}
    \label{fig:state-details}
\end{figure}

This normalization technique is based on the process introduced in \cite{Ross_2013_NormalizedOnlineLearning} and normalizes the data by dividing it by a value during the creation of the state. Let $\vec{X_{t}}$ be the state generated at time step $t$ and, following \cite{Weng_2020_PortfolioTradingSystem}, let $\vec{P_{t}}$ be the closing prices, $\vec{P_{t}}^{(hi)}$ the high prices and $\vec{P_{t}}^{(lo)}$ the low prices of the assets in the portfolio, as represented in Figure \ref{fig:state-details}.

In this work, two types of state normalization are being investigated:

\begin{description}
    \item[By last closing price:] In this approach, all the prices (including high, low and closing prices) are divided by the last closing price. This was the technique applied in the article that introduced the formulation utilized in this paper \cite{Jiang_2017_DeepReinforcementLearning} and its application can be seen in the equation below.
    \begin{align}
    \begin{split}
        \vec{X_{t}} &=  \vec{X_{t}} \oslash \vec{X_{T}}, \\ 
        \vec{X_{t}}^{(hi)} &=  \vec{X_{t}}^{(hi)} \oslash \vec{X_{T}}, \\ 
        \vec{X_{t}}^{(lo)} &=  \vec{X_{t}}^{(lo)} \oslash \vec{X_{T}}.
    \end{split}
    \end{align}
    \item[By last price:] This technique was utilized in \cite{Shi_2019_MultiScaleTemporalFeature} and \cite{Shi_2022_GPMGraphConvolutional} and divides all the prices by its last value. The equation below demonstrates how to apply it:
    \begin{align}
    \begin{split}
        \vec{X_{t}} &=  \vec{X_{t}} \oslash \vec{X_{T}}, \\ 
        \vec{X_{t}}^{(hi)} &=  \vec{X_{t}}^{(hi)} \oslash \vec{X_{T}}^{(hi)}, \\
        \vec{X_{t}}^{(lo)} &=  \vec{X_{t}}^{(lo)} \oslash \vec{X_{T}}^{(lo)}.
    \end{split}
    \end{align}
\end{description}

\begin{figure}[!ht]
    \centering
    \includegraphics[width=0.9\linewidth]{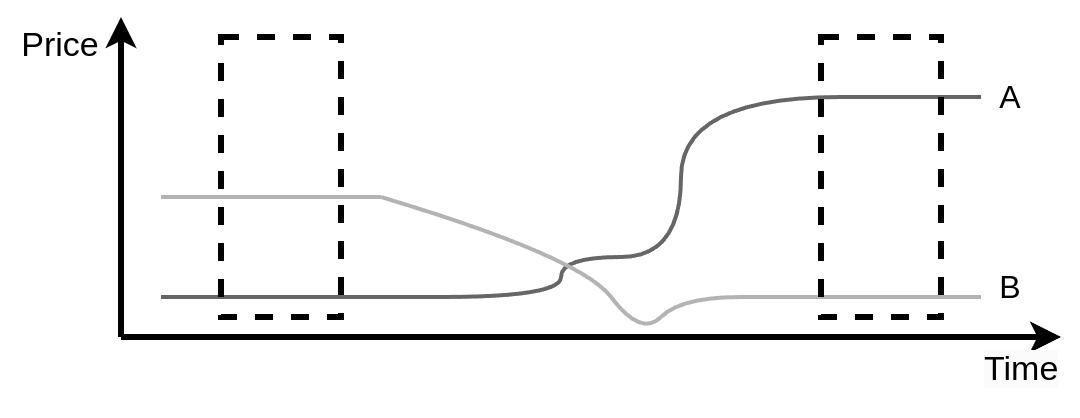}
    \caption{A demonstration of how the agent can lose information by using state normalization. Since the state focuses on the rate of variation of the assets, stock A and  B, in both time windows (denoted by the dotted rectangle), are represented by the same temporal series because they are stable. Therefore, the agent cannot capture that stock B is undervalued.}
    \label{fig:normalization}
\end{figure}

Several variations of this type of normalization have been applied in several other articles \cite{Soleymani_2020_FinancialPortfolioOptimization, Soleymani_2021_DeepGraphConvolutional}, but the core is the same: the state space represents the rate of change of the input time series in the time window, which is adherent to the objective function of Equation \ref{eq:objective-function}. However, this approach makes the agent lose information about the true value of the assets so that it only learns trading strategies that act when the prices are falling or booming. Therefore, this might prevent the agent to, for example, identify that an asset is undervalued and buy it (or sell it, if it is overvalued). Figure \ref{fig:normalization} contains a simple example of how information can be lost using this type of normalization.

\subsection{Data Normalization}

This method, on the other hand, consists of normalizing the time series of the prices of assets independently before using it in the training process. Several time series normalization techniques can be used \cite{Singh_2020_InvestigatingImpactData} but the main focus of this work is to investigate the consequences of simply dividing the whole time series by its maximum absolute value. Therefore, for a temporal sequence $y(t)$, the transformed temporal series can be calculated as
\begin{equation}
    y'(t) = \frac{y(t)}{max(\{y(t): t \in T\})},
\end{equation}
in which $T$ is the set of possible values for $t$.

Differently from the state normalization, this method provides the agent with information about the current value of an asset and it can be used for comparisons with the price history in order to take an action: if the price is close to $0$, the asset can be undervalued, and to buy it might be an interesting investment.

However, since the normalization is first fit to the training data, the price of the assets can achieve values bigger than $1$ in the test period, specially in volatile booming markets. Since the agent was trained with values between $0$ and $1$, unreliable behaviors can be generated by the agent's policy, making it underperform. This issue can be mitigated with the usage of online learning, which is a feature of the PG algorithm introduced in Section \ref{subsec:policy-gradient} and, thus, this normalization is still applicable to the use-case of this work.

\section{Experiments}
\label{sec:experiments}

In order to verify the performance of the agent using different markets and different normalization methods, three portfolios composed of high-volume assets have been constructed:

\begin{enumerate}
    \item A portfolio of stocks of the American market (NYSE) composed of 11 assets: Bank of America Corporation (BAC), Ford Motor Company (F), Regions Financial Corp (RF), Wells Fargo \& Co (WFC), General Electric (GE), Pfizer (PFE), Citigroup (C), AT\&T Inc (T), Marathon Oil Corporation (MRO), United States Steel (X), JPMorgan Chase \& Co (JPM). This is considered a very stable market.
    \item A portfolio composed of 10 Brazilian stocks (IBOVESPA): Vale (VALE3), Petrobrás (PETR4), Itaú Unibanco (ITUB4), Bradesco (BBDC4), Banco do Brasil (BBAS3), Localiza (RENT3), Lojas Renner (LREN3), Petro Rio (PRIO3), WEG (WEGE3), Ambev (ABEV3). In comparison to NYSE, this market is considerably more volatile.
    \item A cryptocurrency portfolio composed of 9 coins: Cardano (ADA), Binance Coin (BNB), Bitcoin (BTC), Bitcoin Gold (BTG), Dogecoin (DOGE), Ethereum (ETH), ChainLink (LINK), TRON (TRX), Ripple (XRP). Between the three markets, this is the one with the biggest volatility.
\end{enumerate}

The historical data is obtained in two different sources. For NYSE and IBOVESPA, the price of the stocks is taken from \textit{Yahoo Finance website}\footnote{Yahoo Finance can be accessed in \url{https://finance.yahoo.com/}.}. The cryptocurrencies' historical prices, on the other hand, were obtained from a dataset available in \textit{Kaggle}\footnote{The dataset can be found in \url{https://www.kaggle.com/datasets/svaningelgem/crypto-currencies-daily-prices}.}. In both cases, the data is composed of daily closing, high and low prices of the components of the portfolio.

\subsection{Procedure}

For each normalization method introduced in Section \ref{sec:normalization}, 50 training and testing processes are run. In the NYSE and IBOVESPA portfolios, the agent is trained in the period between 2011/11/11 and 2019/12/31 and tested in the year of 2020 (from January 1st to December 31st). This test period was chosen because it was challenging for the market since the COVID-19 pandemic has shaken stock prices around the world. Therefore, we are testing one of the worst-case scenarios for the agent. 

Due to the fact that cryptocurrencies are new investments, in the third portfolio, more recent periods were chosen in order to have enough training data: the training period is from 2018/01/01 and 2022/12/31 and the testing one is from 2023/01/01 to 2023/12/31.

For all the experiments, the same hyper-parameters were utilized both during the training and the test process. Table \ref{tab:hyperparameters} summarizes the hyper-parameters applied.

\begin{table}[!ht]
\centering
\caption{Hyper-parameters used in the experiment.}
\label{tab:hyperparameters}
\begin{tabular}{l|c|l}
\textbf{Hyperparameter} & \textbf{Value} & \textbf{Description} \\ 
\hline
Learning rate            & 0.00005        & \begin{tabular}[c]{@{}l@{}}The step size used in the gradient\\ ascent with AdamW optimizer.\end{tabular} \\ 
\hline
Batch size               & 200            & \begin{tabular}[c]{@{}l@{}}Size of the sequence of experiences\\ used in the optimization.\end{tabular} \\ 
\hline
Sample bias              & 0.002          & \begin{tabular}[c]{@{}l@{}}Probability considered in the geometric\\ distribution used when sampling the\\ batches of experiences.\end{tabular} \\ \hline
Steps                    & 300000         & \begin{tabular}[c]{@{}l@{}}Number of training steps used to train\\ the agent. \end{tabular}  \\ 
\hline
Online Steps             & 30             & \begin{tabular}[c]{@{}l@{}}Number of training steps to run after\\ every rebalancing in the testing period.\end{tabular} \\
\hline
Time window              & 50             & \begin{tabular}[c]{@{}l@{}}Size (in steps) of the time window\\ considered in the state representation.\end{tabular} \\ 
\hline
Commission rate          & 0.0025         & \begin{tabular}[c]{@{}l@{}}Rate of comission fee applied when\\ rebalancing the portfolio.\end{tabular} \\ 
\hline
Initial value            & 100000         & \begin{tabular}[c]{@{}l@{}}Initial value of the portfolio in BRL\\ (for IBOVESPA) and USD (for NYSE\\ and cryptocurrencies).\end{tabular} \\
\end{tabular}
\end{table}

\subsection{Performance Metrics}

To quantify the performance of the agent, three performance metrics are utilized in this work:

\begin{description}
    \item[Final Accumulative Portfolio Value (FAPV):] This metric calculates the ratio between the final portfolio value and the initial portfolio value:
    \begin{equation}
        FAPV = \frac{V_{T}^f}{V_{0}},
    \end{equation}
    in which $V_{T}^f$ is the final value of the portfolio and $V_{0}$ is the initial one.
    \item[Maximum Drawdown (MDD) \cite{Magdon-Ismail_2004_MaximumDrawdownBrownian}:] The maximum drawdown is a method to quantify the risk of the portfolio. It represents the biggest loss of the rebalancing during a period of time. The bigger the MDD, the more risky is the trading strategy applied. It can be calculated as:
    \begin{equation}
        MDD =  \max \Bigg( \max_{\ t < \tau \ } \frac{V_{t}^f - V_{\tau}^f}{V_{t}^f} \Bigg).
    \end{equation}
    \item[Sharpe Ratio (SR) \cite{Sharpe_1994_SharpeRatio}:] Finally, the Sharpe ratio is a metric that quantifies the return of the portfolio divided by its standard deviation. The bigger the standard deviation of the returns, the more volatile is the portfolio strategy and, thus, more risky. Therefore, the SR evaluates the amount of profit generated considering the risk of the rebalancing. This metric is given by the following equation:
    \begin{equation}
        SR =  \frac{\mathbb{E}_{t}[\rho_{t} - \rho_{F}]}{\sqrt{var_{t}[\rho_{t} - \rho_{F}]}},
    \end{equation}
    in which $\rho_{t} = V_{t}^f / V_{t-1}^f$ is the return ratio of the portfolio in step $t$ and $\rho_{F}$ is the return ratio of the risk-free investment (which is set as zero in this work).
\end{description}

A portfolio strategy is considered good when it achieves FAPV as bigger than $1$ as possible (indicating it made profit), MDD close to $0$ (demonstrating a low risk) and high positive SR.

\section{Results}
\label{sec:results}

The experiments\footnote{The code utilized can be found in \url{https://github.com/CaioSBC/norm_portfolio}.} described in Section \ref{sec:experiments} were conducted using \textit{RLPortfolio} \cite{Costa_2025_RLPortfolioReinforcementLearning}, an open source library for training portfolio optimization agents using reinforcement learning. The results are discussed below.

\subsection{In the American Market}
\label{subsec:america}

Table \ref{tab:results-nyse} demonstrates the aggregated results of the 50 runs conducted. It can be noted that using data normalization produces the best outcome in terms of FAPV and SR, showing that the agent is benefiting from the information about the true value of the assets. With respect to the MDD metric, the normalization by the last price achieves the best performance but, considering its margin of error, it is very similar to the data normalization. All these results indicate that to normalize the date before training is more effective in this market.

\begin{table}[!ht]
\centering
\caption{Mean results of the NYSE portfolio.}
\label{tab:results-nyse}
\begin{tabular}{c|c|c|c}
\textbf{\begin{tabular}[c]{@{}c@{}}Normalization\\ Method\end{tabular}} & \textbf{FAPV}        & \textbf{MDD}         & \textbf{SR}            \\ \hline
\textit{\begin{tabular}[c]{@{}c@{}}Last Closing\\ Price\end{tabular}}   & 0.76 ± 0.06          & 0.60 ± 0.04          & -0.005 ± 0.006         \\ \hline
\textit{\begin{tabular}[c]{@{}c@{}}Last \\ Price\end{tabular}}          & 1.05 ± 0.10          & \textbf{0.52 ± 0.04} & 0.018 ± 0.010          \\ \hline
\textit{\begin{tabular}[c]{@{}c@{}}Data\\ Normalization\end{tabular}}   & \textbf{1.19 ± 0.08} & 0.55 ± 0.02          & \textbf{0.033 ± 0.006}
\end{tabular}
\end{table}

It is interesting to highlight the poor performance of the normalization by the last closing price, which can be seen in details in Figure \ref{tab:results-nyse}. The majority of the runs using this technique produces suboptimal policies that make the agent lose money ($FAPV < 1$). The other state normalization strategy, on the other hand, generates a more heterogeneous distribution, which makes it the most unreliable: it can produce agents that have considerable losses but also ones that multiply the portfolio value by $2$. 

\begin{figure}[!ht]
    \centering
    \includegraphics[width=\linewidth]{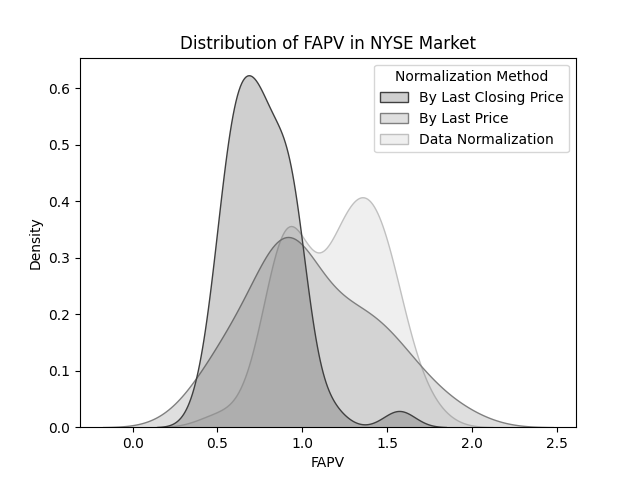}
    \caption{Graph demonstrating the approximate distribution of the results in the NYSE market.}
    \label{fig:nyse-distribution}
\end{figure}

Finally, Figure \ref{fig:nyse-distribution} also demonstrates that the data normalization method appears to produce, in most cases, two policies: one that makes the agent preserve its value and another that makes it multiply its value by $1.5$. Additionally, the distribution has a smaller variance, indicating that this normalization technique is more reliable.

\subsection{In the Brazilian Market}
\label{subsec:brazil}

In the Brazilian market, it is observed a pattern similar to the one noticed in the American market but, since the Brazilian market is more volatile, the differences in the metrics are more noticeable when comparing. As it can be seen in Table \ref{tab:results-ibovespa}, the data normalization method achieves the higher performance in both FAPV and SR metrics, indicating that it is the most suitable for this market. The main downside of this method, however, can be observed in the MDD metric, which reveals that it leads to more risky agents.

\begin{table}[!ht]
\centering
\caption{Aggregated results of the IBOVESPA portfolio.}
\label{tab:results-ibovespa}
\begin{tabular}{c|c|c|c}
\textbf{\begin{tabular}[c]{@{}c@{}}Normalization\\ Method\end{tabular}} & \textbf{FAPV}        & \textbf{MDD}         & \textbf{SR}            \\ \hline
\textit{\begin{tabular}[c]{@{}c@{}}Last Closing\\ Price\end{tabular}}   & 1.48 ± 0.16          & 0.57 ± 0.02          & 0.051 ± 0.010         \\ \hline
\textit{\begin{tabular}[c]{@{}c@{}}Last \\ Price\end{tabular}}          & 1.26 ± 0.16          & \textbf{0.55 ± 0.02} & 0.040 ± 0.008          \\ \hline
\textit{\begin{tabular}[c]{@{}c@{}}Data\\ Normalization\end{tabular}}   & \textbf{1.79 ± 0.10} & 0.7 ± 0.04           & \textbf{0.069 ± 0.006}
\end{tabular}
\end{table}

\begin{figure}[!ht]
    \centering
    \includegraphics[width=\linewidth]{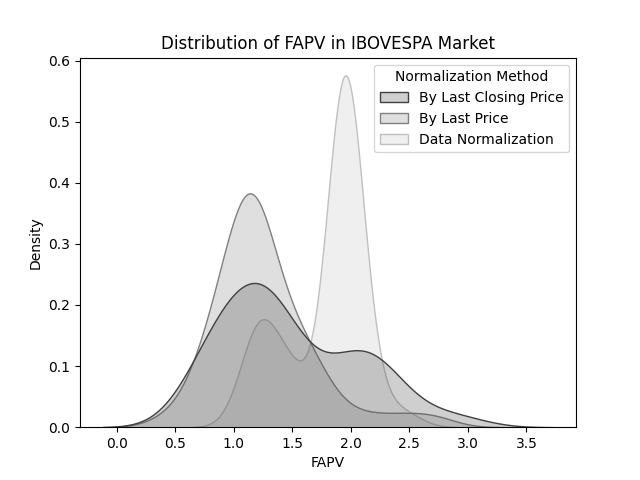}
    \caption{Graph demonstrating the approximate distribution of the results in the IBOVESPA market.}
    \label{fig:ibovespa-distribution}
\end{figure}

Figure \ref{fig:ibovespa-distribution} demonstrates the approximate distribution of the FAPV metric along the runs and it is noticeable that, just like in the American market, the data normalization is more consistent. But it can also be noticed that the behavior of both state normalizations have switched: this time, the normalization by last closing price has the bigger variance instead of the one by last price. This fact evidences that, when using a state normalization method, it is important to choose it wisely and test several possibilities because, depending on the market, one method behaves considerably different than the other.

\subsection{In the Cryptocurrency Market}
\label{subsec:crypto}

Curiously, the cryptocurrency market, the one utilized in the paper that introduced the formulation adopted in this work \cite{Jiang_2017_DeepReinforcementLearning}, is the one which benefited the most by the data normalization method. As it can be seen in Table \ref{tab:results-crypto}, the performance of the agent when using this method largely outperforms the other ones in all considered metrics.

\begin{table}[!ht]
\centering
\caption{Aggregated results of the cryptocurrency portfolio.}
\label{tab:results-crypto}
\begin{tabular}{c|c|c|c}
\textbf{\begin{tabular}[c]{@{}c@{}}Normalization\\ Method\end{tabular}} & \textbf{FAPV}        & \textbf{MDD}         & \textbf{SR}            \\ \hline
\textit{\begin{tabular}[c]{@{}c@{}}Last Closing\\ Price\end{tabular}}   & 0.78 ± 0.10          & 0.54 ± 0.04          & -0.025 ± 0.012         \\ \hline
\textit{\begin{tabular}[c]{@{}c@{}}Last \\ Price\end{tabular}}          & 0.70 ± 0.10          & 0.56 ± 0.04          & -0.034 ± 0.012          \\ \hline
\textit{\begin{tabular}[c]{@{}c@{}}Data\\ Normalization\end{tabular}}   & \textbf{1.78 ± 0.04} & \textbf{0.44 ± 0.02} & \textbf{0.059 ± 0.002}
\end{tabular}
\end{table}

\begin{figure}[!ht]
    \centering
    \includegraphics[width=\linewidth]{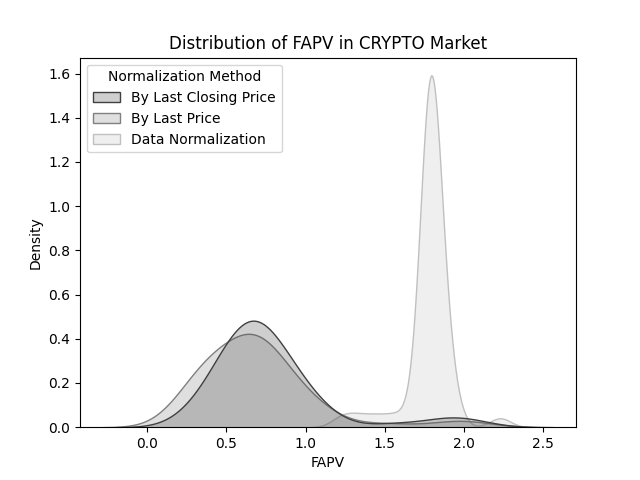}
    \caption{Graph demonstrating the approximate distribution of the results in the cryptocurrencies market.}
    \label{fig:crypto-distribution}
\end{figure}

It can also be noticed that the state normalization technique is poorly effective, since the majority of the test runs yields losses in the portfolio value and, thus, produces an agent that loses money. As demonstrated in Figure \ref{fig:crypto-distribution}, the two state normalization methods generate similar distributions, meaning that the choice between them is considerably irrelevant.

It is interesting to discuss the possible issues that made the cryptocurrency portfolio underperform in this experiments while the portfolio used in \cite{Jiang_2017_DeepReinforcementLearning} achieved great results. The first issue is the fact that the portfolios utilized are not the same: since the original article does not provide the list of assets in their portfolio, this paper simply constructed a portfolio with high volume coins in order to adhere to the formulation presented in \ref{sec:formulation}. Additionally, this article makes use of daily historical prices, which can be easily found in several databases on-line, while \cite{Jiang_2017_DeepReinforcementLearning} generates a portfolio that rebalances its assets every 30 minutes. Finally, since the input data of this work has smaller granularity, more data points are necessary and, thus, a bigger time period is used to train and test the data. The time series utilized are closer to the present time, in which the cryptocurrency market is a little more stable than in \cite{Jiang_2017_DeepReinforcementLearning}: in the latter, the agent is trained and tested in periods of the years of 2017 and 2018, when several cryptocurrencies were surging and booming while, in this work, the test period is the year of 2023.

The possible influence of these changes in the input data underlines that the approach introduced in \cite{Jiang_2017_DeepReinforcementLearning} is not as general as it seems and that optimizations in the hyperparameters, normalization methods and even changes in the objective function might be necessary when applied to different markets.

\subsection{Analysis of the Results}

As it can be seen in Sections \ref{subsec:america}, \ref{subsec:brazil} and \ref{subsec:crypto}, the positive effects of the data normalization are bigger the bigger the volatility of the target market. From Figure \ref{fig:nyse-distribution} to Figure \ref{fig:crypto-distribution}, it is noticeable that the overlap of the distributions of the FAPV gets smaller, showing that the volatility of the market actually degrades the performance of the agent under the state normalization.

\begin{figure}[!ht]
    \centering
    \includegraphics[width=1\linewidth]{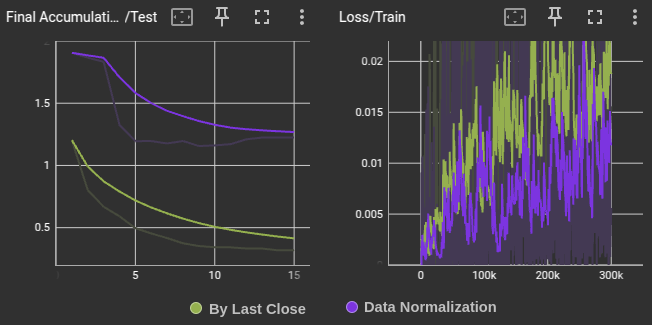}
    \caption{Example of logs of runs in which overfit occurs. In the left, the graph shows the FAPV over 15 performance tests conducted in the test period while the agent was being trained. In the right, the loss function in the training process is plotted for every step. The data is smoothed to outline trends.}
    \label{fig:tensorboard}
\end{figure}

Additionally, it is observed in the training logs that the state normalization is more susceptible to overfitting: in the cryptocurrency market, for example, the majority of the runs produce agents whose performance deteriorates over the training process, as seen in Figure \ref{fig:tensorboard} (generated with \textit{Tensorboard} \cite{Abadi_2016_TensorFlowSystemLargescale}), which demonstrates the behavior of the agent in the test period during the training process in the worst case scenario (i.e. the run whose training process generated the most overfitting). Such behavior also happens when using data normalization, but the agent gets stuck in a much better local minimum as seen in Figure \ref{tab:results-ibovespa}.

The bigger tendency of overfitting in the state normalization can be justified by the fact that it simplifies the state space more than the data normalization. While the latter keeps information about the true value of the assets, the other one reduces the observations of the agent to price variations so that different situations of the market can be considered the same, as explained in Section \ref{sec:normalization}. Therefore, different moments of the market in the testing period generate states that are approximately equal to states in the training period, so that the agent's deterministic policy generates suboptimal actions. The more the training process proceeds, the more the agent specializes in the training period, generating worse actions when testing.

Finally, it is important to highlight that, even though the data normalization is more reliable statistically, all the normalization methods eventually produced good agents. Therefore, it is possible to run several training processes, choose the best agent considering its performance in a validation period and, then, test it. However, the more unreliable the normalization method, the more difficult it is to find a good agent, the reason why the data normalization technique is favored. Table \ref{tab:results-max} contains the maximum FAPV achieved by the experiments described in Section \ref{sec:experiments} and it can be noticed that there is no clear pattern to identify which method yields the best result across multiple markets.

\begin{table}[!ht]
\centering
\caption{Max FAPV for each market and normalization method.}
\label{tab:results-max}
\begin{tabular}{c|c|c|c}
\textbf{\begin{tabular}[c]{@{}c@{}}Market of\\ Portfolio\end{tabular}}  & \textbf{NYSE}        & \textbf{IBOVESPA}    & \textbf{CRYPTO}        \\ \hline
\textit{\begin{tabular}[c]{@{}c@{}}Last Closing\\ Price\end{tabular}}   & 1.57                 & \textbf{2.94}        & 2.06                   \\ \hline
\textit{\begin{tabular}[c]{@{}c@{}}Last\\ Price\end{tabular}}           & \textbf{1.96}        & 2.71                 & 2.06                   \\ \hline
\textit{\begin{tabular}[c]{@{}c@{}}Data\\ Normalization\end{tabular}}   & 1.74                 & 2.47                 & \textbf{2.23}
\end{tabular}
\end{table}

\section{Conclusion}
\label{sec:conclusion}

This article demonstrates empirically that, in the portfolio optimization domain, to normalize the input data instead of the RL agent's state improves the performance of the agent not only in the stock market but also in the cryptocurrency market. The results of the experiments showed that the more volatile a market is, the more it benefits from the usage of the data normalization since the agent does not suffer from the information loss inherent to the state normalization methods.

Additionally, even though the data normalization technique has consistently outperformed the state normalization, it is concluded that the latter can produce effective agents, but its high variability can make the process considerably slower since the agent tends to converge to worse suboptimal policies. It was also shown that, when using state normalization, the best technique can change depending on the market, so it is better to treat it as a hyperparameter and optimize it carefully.

Finally, the good results of the data normalization indicates that the online learning is able to counterweight the fact that this type of normalization can produce agents that are not prepared to handle ranges of input data that surpasses the training data. Future works can further investigate this balance in order to understand the relationship between online learning rates and online training steps in the performance of an agent that uses this type of normalization.



\bibliographystyle{IEEEtran}
\bibliography{references}

\end{document}